# Exploring the role of *nonlocal* Coulomb interactions in perovskite transition metal oxides

Indukuru Ramesh Reddy[1], Chang-Jong Kang[2], Sooran Kim[3] & Bongjae Kim[1*]

[1]*Department of Physics, Kyungpook National University, Daegu 41566, Korea*
[2]*Department of Physics, Chungnam National University, Daejeon 34134, Korea*
[3]*KNU G-LAMP Project Group, KNU Institute of Basic Sciences, Department of Physics Education, Kyungpook National University, Daegu 41566, Korea*

*E-mail: bongjae@knu.ac.kr

Employing the density functional theory incorporating on-site and inter-site Coulomb interactions (DFT+$U$+$V$), we have investigated the role of the nonlocal interactions on the electronic structures of the transition metal oxide perovskites. Using constrained random phase approximation calculations, we derived screened Coulomb interaction parameters and revealed a competition between localization and screening effects, which results in nonmonotonic behavior with $d$-orbital occupation. We highlight the significant role and nonlocality of inter-site Coulomb interactions, $V$, comparable in magnitude to the local interaction, $U$. Our DFT+$U$+$V$ results exemplarily show the representative band renormalization, and deviations from ideal extended Hubbard models due to increased hybridization between transition metal $d$ and oxygen $p$ orbitals as occupation increases. We further demonstrate that the inclusion of the inter-site $V$ is essential for accurately reproducing the experimental magnetic order in transition metal oxides.

**Introduction:**

Hubbard model attempt to explains the electronic structures of a correlated system employing a Hamiltonian with minimal parameters including on-site Coulomb interaction, $U$, and hopping parameter, $t$[1]. While the Hamiltonian itself is simple, it explains the essential features in a material with strong electronic Coulomb interactions. The realistic many-body electron systems are projected or simplified to a low-energy Hubbard model Hamiltonian, which is then solved (near-) exactly employing various numerical approaches, and one can readily obtain a rich phase diagram within the parameter space. For transition metal oxides (TMOs), a representative class of strongly correlated systems, emergent phases such as magnetism, Mott electronic transition, and superconductivities are well-described based on the Hubbard model[2–4]. While the recent developments in the numerical and computational approaches with various lattice-types and bands are advancing the field and offers new insights from simple model perspectives[5,6], the connection to the realistic system is often not transparent. This is because such methods require projection from the many-body electrons to a few basis, which leads to the oversimplification of the real situation and readily suffer finite-site effect problems.

Density functional theory (DFT) has offered the accessible route to obtaining the ground state properties of a realistic system. In DFT, one directly solves the many-body electron Hamiltonian of a solid system by mapping it onto a non-interacting electron Hamiltonian in an effective potential[7]. However, the inability of DFT in the description of the many-body on-site interactions limit its use, and its direct application to correlated systems, representatively, transition metal oxides, often failed. Integrating the many-body model methods into DFT, in the form of DFT+$U$ and DFT+DMFT, better accounted for the Coulomb correlation among the electrons and has been a great success in the theoretical condensed matter physics community[8–11]. Replacing the mean-field part of the Coulomb on-site correlation with a Hubbard-type term dramatically improved the description of the electronic structure of the correlated system, and one can obtain better description of the key experimental features such as the band gap, optical properties, and photoemission spectra in TMOs[12,13].

Despite the triumph of the DFT with *local* (on-site) Coulomb interaction, $U$, recent reports show the incapability of the local theory in understanding the physics of the correlated systems[14,15] and finds the importance of the *nonlocal* (inter-site) Coulomb interactions[14,16–19]. In fact, among the model communities, this has been noticed earlier[18,20], and the explicit role of the inter-site, or next nearest neighbor, Coulomb interaction ($V$) on the electronic structures has been intensively discussed[21–24]. Especially, in various two-dimensional (2D) lattices, the inter-site interaction critically decides the charge density waves phases[25–28] and selectively stabilizes the different superconducting order parameters[29–33]. However, there are no corresponding studies in the three-dimensional (3D) lattices, which is more akin to the real material cases.

One may simply expect in 3D systems, the strong Thomas-Fermi screening effectively curtails the nonlocal interactions and the interactions decay exponentially upon distance[34]. In reality, the nonlocal screening extends over the interatomic distances, and 3D systems show sizable inter-site Coulomb interactions as in the 2D cases[35]. Typically, inter-site interaction is known to renormalize the on-site interaction from effectively screening[19,33]. Also, the nonlocal Coulomb interaction renormalizes bandwidth in opposite way compared to on-site interactions[23,24]. Such effects are meticulously studied from the model approaches, usually for the partially-filled one-band case, but, again, extensions to the real materials, where the various occupations are available and multi-orbitals are involved, are rare. Recent development of the GW+DMFT method offer promising practical applications but the computational cost of the GW as well as the DMFT prohibits their wide utilization to the diverse materials[13,36–38].

In this work, we systematically analyze the roles of the inter-site Coulomb interactions in the representative perovskite TMOs with the form of Sr$M$O$_3$, where $M$ = Ti($d^0$), V($d^1$), Cr($d^2$), Mn($d^3$), Fe($d^4$), and Co($d^5$). We first extract on-site and inter-site Coulomb interaction parameters from the first-principles calculation using the constrained random-phase approximation (cRPA) scheme. We show that the competition between the electron localization and the screening from the hybridized orbitals primarily determines the strengths of the Coulomb interaction parameters, which does not exhibit the monotonic behaviors with respect to $d$-electron occupation. We demonstrate that the Coulomb interactions do not decay exponentially within the considered crystal lattices, indicating that the sizable nonlocal interactions are in effect. Utilizing the calculated on-site and inter-site interaction parameters, we perform the DFT+$U$+$V$ calculations for the Sr$M$O$_3$ series. The electronic structures are systematically investigated, and the peculiar role of inter-site Coulomb interaction, $V$, which often exhibits contrasting behaviors upon occupation, is addressed. Furthermore, we show that the inclusion of $V$ correctly reproduces the experimentally observed magnetic configurations compared to the DFT+$U$ results.

## Results and Discussions

### 1. Local and nonlocal Coulomb interactions

To investigate the quantitative local and nonlocal Coulomb interaction parameters and their explicit roles, we considered the representative 3d transition metal oxides in the perovskites form SrMO$_3$ (M = Ti, V, Cr, Mn, Fe, and Co) as shown in Fig. 1. We have systematically changed the nominal occupation of the d-orbitals from $d^0$ (SrTiO$_3$) to $d^5$ (SrCoO$_3$). To account for the inter-site Coulomb interactions, we have constructed √2×√2×2 supercell, including 4 formula units from the 5-atom cubic unit-cell (Fig. 1)[39]. The experimental lattice parameters of the cubic perovskite SrMO$_3$ used in our study are given in Supplementary Table 1.

First, we have investigated the sizes of the Coulomb interaction parameters in the representative perovskite TMOs with the form of SrMO$_3$ (M = Ti, V, Cr, Mn, Fe, and Co). For simplicity, the nonmagnetic phase was chosen to focus solely on the role of local and nonlocal Coulomb interactions. Note that local and nonlocal Coulomb interactions could induce magnetism in some systems, which makes interpretations more complicated. Here, we would like to focus on the generic roles of local and nonlocal Coulomb interactions clearly and straightforwardly in these systems. The importance of the nonlocality of the Coulomb interaction will be discussed later.

We start with the partial density of states (PDOSs) of target systems computed with standard DFT calculations without any U and V parameters, and the results are shown in Figure 2. As the atomic number of the transition metal (TM) increases, more electrons are occupied in the d-orbitals. Starting from the SrTiO$_3$, a gapped $d^0$-system, electrons are occupied in the $t_{2g}$ manifolds from SrVO$_3$ ($d^1$) to SrCoO$_3$ ($d^5$). As the electrons are occupied, the $t_{2g}$ bands narrow, indicating the increased localization of the d-orbitals. This naturally leads to the enhanced electron correlations. The O-p bands, which are located below the d-bands for the unoccupied ($d^0$) case (Fig. 2(a)), progressively move to the higher energy as the occupation increases and the overlap with the TM-d is enhanced. This shows the hybridization between TM-d and O-p bands increases for later TMOs, and the role of the d-p channel in the screening process becomes more important. Specifically, the $e_g$-orbitals hybridize much strongly with the O-p orbitals, hence the PDOSs of $e_g$-bands moves closer to the O-p ones as more electrons are occupied.

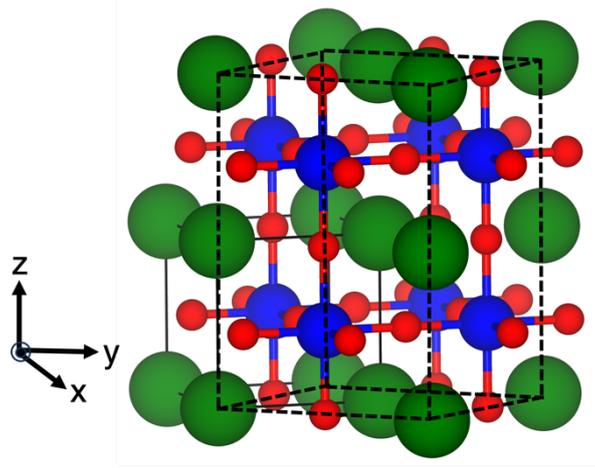

**Fig. 1**. Crystal structure of SrMO$_3$ perovskite. Solid and dashed lines represent the single unit cell and √2×√2×2 supercell, respectively. Green, blue, and red spheres indicate Sr, M (M = Ti – Co), and O atoms, respectively. The axes correspond to the single unit cell.

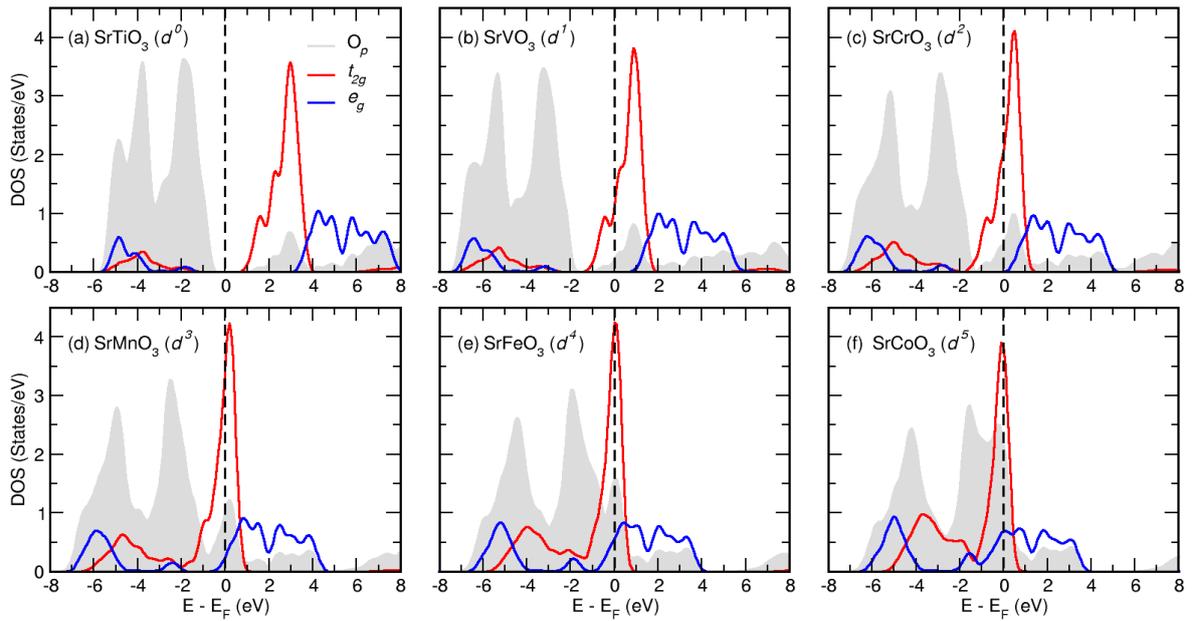

**Fig. 2**. (**a-f**) Partial density of states (PDOSs) of SrMO$_3$ (M = Ti - Co) for a single unit cell. The $t_{2g}$ and $e_g$ orbitals are denoted by red and blue solid lines, respectively. Grey shading represents O-$p$ orbitals. The Fermi level (E$_F$) is set at 0 eV.

Two different schemes were used for constraint RPA (cRPA) calculations to determine the Coulomb interaction parameters: the *d-dp* and the *d*-d scheme. In the *d-dp* scheme, Wannier functions are constructed within the entire TM-*d* and O-*p* energy range, and screened interactions are calculated by excluding the *d*-bands. In the *d-d* scheme, on the other hand, *d*- only Wannier orbitals are constructed, and the excluded polarizability is only for the internal *d*-bands (see the Methodology section below for details).

Figures 3(a)-(d) show the calculated cRPA Hubbard parameters: bare and screened Coulomb / Hund exchange interaction values, $U^{bare}$ / $J^{bare}$, and $U$ / $J$ within the *d-dp* model. As anticipated, when the screening effects are not considered $U^{bare}$ increases with the number of electrons[40]. This trend can be directly attributed to the electron localization in the *d*-orbitals, which is well represented in the Wannier function spreads (Fig. 3(e)). The spread of the Wannier functions, which provides a measure of orbital localization, clearly decreases upon occupations, and is directly reflected in the $U^{bare}$ and $J^{bare}$ values.

Now we consider the screened Coulomb interaction parameters, $U$ and $J$, which are often employed in the construction of the low-energy effective Hubbard models. The first notable feature is that the screened Coulomb interaction does not follow the trend of $U^{bare}$ (see Fig. 3(c)). Opposite to the bare values, the screened $U$ parameter decreases upon occupation except for SrMnO$_3$ and SrFeO$_3$ case, which is quite counterintuitive considering the increased localization upon the occupation[41]. This renormalization is mainly from the electronic screening from the various sources[40,42,43], and for our systems *d-d* and *d-p* screenings are key contributors.

As we noticed from the PDOSs, the energy separation between O-*p* and TM-*d* orbitals are continuously reduced, suggesting enhanced screening from the *d-p* channel. On top of expected *d-d* screenings, this *d-p* channel strongly contributes to the overall screening process, which even reverses the overall trends of the screened Coulomb interactions. As quantified in the $U/U^{bare}$ ratio in Fig. 3(f), the strength of the screening is strongly intensified as the occupation increases: $U/U^{bare}$ is 0.282 for SrTiO$_3$ which decreases to 0.153 for SrCoO$_3$. Furthermore, the energy separation between $t_{2g}$ and $e_g$ levels decreases upon electron occupation, as evident from the changes in the crystal field. This indicates increased screening from the RPA process through particle-hole pair bubbles, and contributes for the increased screening upon the occupation[40]. On the other hand, screened Hund interaction, $J$, exhibits much weaker dependence upon screening, and follows the similar trends of $J^{bare}$. The calculated unscreened and screened cRPA Hubbard parameters within the *d-dp* and *d-d* models are tabulated in Supplementary Table 2. Orbital-resolved $U$ and $V$ parameters for $t_{2g}$ and $e_g$ orbitals obtained from the *d-dp* model are given in Supplementary Table 3, and the related discussion is provided in Supplementary Note 3.

The importance of *d-p* screening can be further confirmed by comparing it with cRPA calculation using the *d-d* model. Here, the both screened and bare $U$ values decrease monotonically upon occupation as shown in Fig. 3(c). This indicates the shortage of Wannier basis in the description of the system. The spread in the *d*-orbitals is much larger for the *d*-only basis compared to the *d-p* basis, and there is a sudden jump of the spread curve for the later TMOs of $SrFeO_3$ and $SrCoO_3$ (see blue dashed line in Fig. 3(e)), demonstrating the incompleteness of the basis orbitals. Despite the apparent localization of the *d*-orbitals upon occupation (Fig. 2 and Supplementary Fig. 1), the spread of the *d*-like Wannier orbitals increases for *d-d* model (Fig. 3(e)). This unphysical behavior is directly reflected in the $U^{bare}$ values in the *d-d* model – the overall increase in the spread can explain the decrease in the $U^{bare}$ upon occupancy. The screening does not change the overall tendency of the interaction parameters, and the screened $U$ follows the same trend within the *d-d* model. Considering the strong the *dp*-hybridization for TMOs, neglecting this would not properly capture the overall screening process of the system. This is particularly true for the late TMOs such as $SrCoO_3$ and $SrFeO_3$, where the hybridization is significantly strong, as shown in the orbital-resolved band structures (Supplementary Fig. 2).

Nonmonotonic behaviors of screened $U$ observed for the *d-dp* model demonstrate the complexity of actual physics, and the *d-p* channel playing an important role. Wannier localization increases the interaction parameters, while screening reduces them. The latter effect is more dominant in the *d-dp* model, resulting in an overall decreasing tendency for screened $U$ upon occupancy. However, in the case of $SrFeO_3$, the screening is not strong enough, leading to an overturn behavior appears that causes nonmonotonicity[44]. This active involvement of the *p*-orbital suggests that the physics of the highly-occupied *d*-system largely deviates from the extended Hubbard model, where only the correlated *d*-orbitals are considered. Comparison with the model will follow later.

Now we turn to the inter-site Coulomb interactions (*V*) obtained from the a $\sqrt{2}\times\sqrt{2}\times2$ supercell[45]. Note that the *V* parameter exhibits a trend similar to that of *U* (see Supplementary Fig. 4). Figure 4 plots the real-space profile of the Coulomb interaction for on-site, first nearest-neighbor, and second nearest-neighbor. The values are renormalized to the on-site Coulomb values. The exact interaction parameters are shown in Table 1, and we have provided the range of inter-site Coulomb interactions here, which vary for different neighbor sites as the screening varies. While we have three sets of screened interaction parameters, on-site $U$, and nearest and next-nearest neighbor $V$, we clearly observe that the interaction does not decay rapidly with distance[35,46], which is very different from the textbook charged-impurity model with a form of $U(r) = \frac{e^2}{r}\exp(-\lambda r)$[34]. Here $\lambda$ is a screening radius[31]. For the bare interaction case, the decay is still strong and nearest-neighbor interaction, $V$, is about 20% for the local one. The real-space decay of the $U^{bare}$, influenced by the spatial extent of the Wannier functions, $\delta$, can be described using local screening model, i.e., the Ohno potential[47,48]:

$$\frac{U^{bare}(r)}{U^{bare}(r=0)} = \frac{1}{\sqrt{\frac{r^2}{\delta^2}+1}}$$

And our fit of the Coulomb parameter, with $\delta$ = 0.76 Å, well describes the screening decay behavior of $U^{bare}$ and is consistent with previous study[35].

Interestingly, the screened Coulomb interaction profile clearly shows the nonlocal feature. The decaying behavior is much softer, and the $V$ value is about 37% of the local interaction, $U$ (Table 1). We can fit the interaction parameters from the $d$-$dp$ model to the Ohno-Resta potential, and found an effective ionic radius $R$ of 6.35 Å[35,49]. This effective ionic radius represents the boundary beyond which the interaction behaves as local. In our representative perovskites, where the typical lattice parameter is less than 4 Å, the first and second nearest-neighbor TM sites falls within the effective ionic radius, placing the interaction parameters in the nonlocal regime. Our systematic analysis strongly demonstrates the nonnegligible size of the nonlocal Coulomb interactions, which has been noticed in a few systems[35,43,50,51], and, as we will demonstrate below, has a strong implication on the computational description of the correlated systems.

**Table 1.** On-site ($U$) and range of inter-site ($V$) Hubbard values (in eV) obtained from the *d-dp* model for Sr$M$O$_3$ perovskites in a $\sqrt{2}\times\sqrt{2}\times 2$ supercell. The ranges of $V$ values relate to different nearest-neighbor metal atoms, since screening vary with the metal-metal distance.

|   | Ti | V | Cr | Mn | Fe | Co |
|---|---|---|---|---|---|---|
| $U$ | 5.48 | 4.64 | 4.10 | 4.27 | 5.15 | 4.10 |
| $V$ | 1.48-1.72 | 1.29-1.39 | 1.29-1.31 | 1.30-1.32 | 1.38-1.75 | 1.28-1.45 |

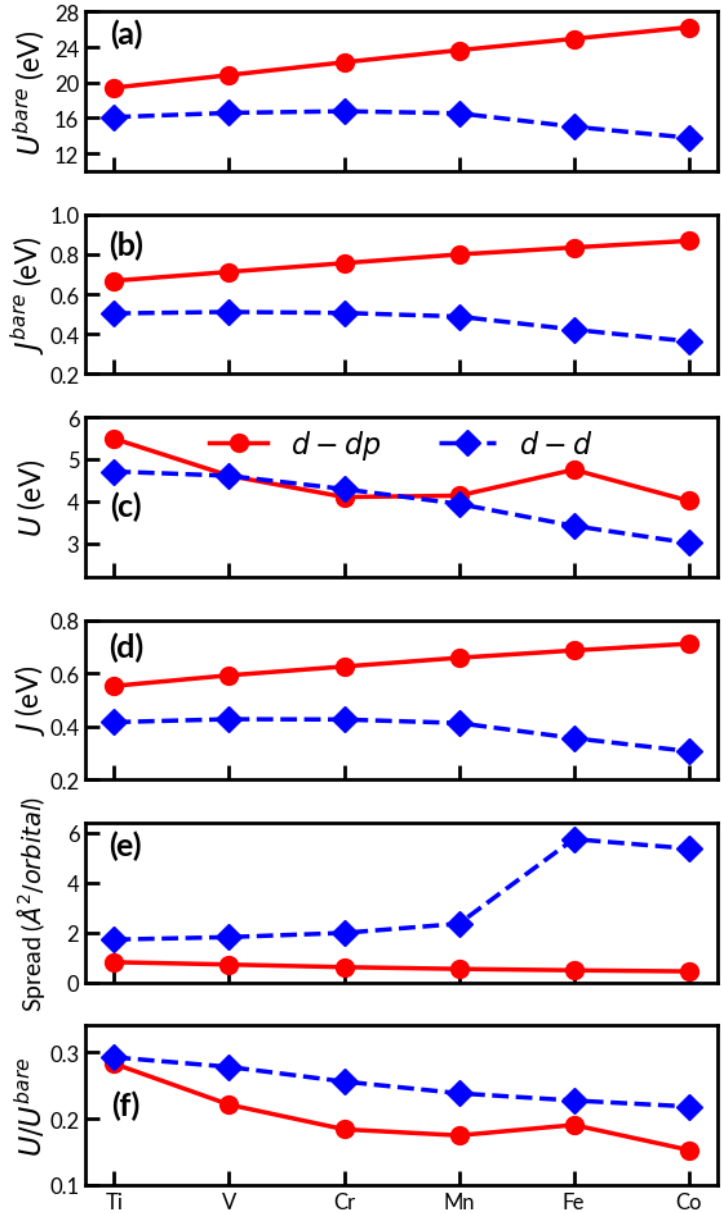

**Fig. 3**. Evolution of the cRPA Hubbard parameters ($U^{bare}$, $J^{bare}$, $U$, $J$) and the spread of Wannier function. The red solid (blue dashed) lines with filled circles (diamonds) represent the values obtained from *d-dp* (*d-d*) model.

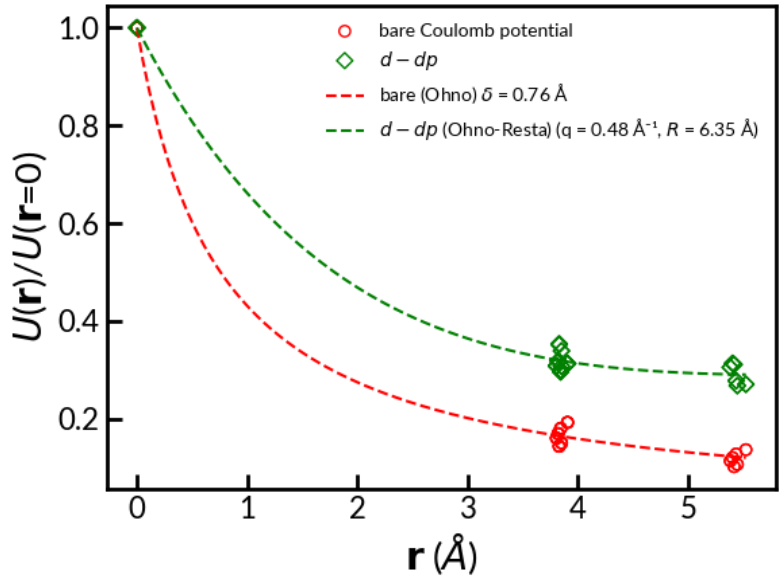

**Fig. 4**. The scaling behavior of the bare (red circles) and screened (green diamonds) Coulomb interaction in Sr$M$O$_3$ perovskites. The fits corresponding to the screening models (see text for details) are represented with dashed lines.

## 2. Role of the inter-site interactions

### 2-1. Electronic structures

Having quantitatively analysed the local and nonlocal Coulomb interaction values, we are now to investigate the realistic electronic structures of representative Sr$M$O$_3$ systems. To understand the role of the on-site and inter-site interaction in target materials, we compared the electronic structures obtained from DFT, DFT+$U$ and DFT+$U$+$V$ calculations, where the latter is an extension of well-known DFT+$U$ approach with further inclusion of inter-site interaction, $V$. There have been many discussions on the implementation and applications of DFT+$U$+$V$ formalism[52–54]. In most of the studies, additional interactions $V$ are those for O-$p$ orbitals or the one between TM-$d$ and O-$p$ orbitals[55–57], not solely from the correlated TM-$d$ manifolds. In our studies, however, in line with the low-energy model approaches, Hubbard-type interactions are accommodated only at the correlated orbitals, *i.e.* on-site and inter-site TM-$d$ orbitals, and this directly corresponds to the extended Hubbard models.

The inter-site Coulomb interaction is known to renormalize *both* local interactions and bandwidth[23,24,58]. The reduction of the effective local interactions by nonlocal interaction has been studied within half-filled Hubbard model[59,60], and this renormalization successively explained the electronic structure of a few systems[21,61]. On top of the band renormalization of the local many-body interactions, the nonlocal interaction is known to renormalize the active bands in the opposite way, that is, it widens the bandwidth. Such modification and screening effects on the local interaction and band renormalization are considered to be the hallmarks of the inter-site interactions[23,24]. In reality, however, many orbitals are involved and the filling of the orbitals plays an important role. This makes the analogy to the extended Hubbard model not very straightforward as we will discuss below.

Figure 5 shows the DFT, DFT+$U$ and DFT+$U$+$V$ PDOSs of the whole Sr$M$O$_3$ ($M$ = Ti, V, Cr, Mn, Fe, and Co) series utilizing our calculated $U$ and $V$ interaction parameters. For SrTiO$_3$, there is no electrons occupied in $d$-orbital ($d^0$) and DFT+$U$ simply pushes unoccupied $d$-bands to the higher energy compared to the DFT results. In this case, inclusion of inter-site interaction $V$ pushes further up the lowest unoccupied states, increasing the band gap, and acts as *additional U*. This is also reflected in the bandwidths such that adding onsite $U$ decreases the bandwidth and inter-site $V$ further decreases one. The band gap increases from DFT value of 1.88 to 2.51 eV with $U$, and then to 2.74 eV with further $V$. We clearly see the inter-site interaction $V$ further correct the size of the gap from DFT+$U$, and moves closer to the experimental value of 3.25 eV[62]. Note that the GW calculation, where the nonlocality is fully treated, overestimates the unoccupied levels, hence, the band gap by ~0.6 eV[63,64]. DFT+$U$+$V$ provides the moderate correction from the DFT and DFT+$U$ result. Here, we further note that increasing $V$ results the experimental band gap but without altering the overall shape of the electronic structure[65]. For SrTiO$_3$ with unfilled $d$-shell, inter-site $V$ increases the effect of local interaction $U$ in general. This is different from the other set of systems, which has a partially-filled correlated orbitals.

The physics of $d^1$ and $d^2$ systems, SrVO$_3$ and SrCrO$_3$ can be directly compared to that of the partially-filled Hubbard model as the position of O-$p$ orbitals is well-below Fermi level and hybridization with the correlated TM-$d$ orbitals is small. Fig. 5(b) and (c) show the TM-$d$ DOS hardly changed from DFT upon the inclusion of local $U$, especially at around the Fermi energy. This is expected because many-body band renormalization is not captured within the mean-field DFT calculations, and partially filled $d$-orbitals of transition metals are not significantly altered by static $U$. However, when inter-site interaction $V$ is further included, we can clearly see the differences. Widening of the $d$-bandwidths, which is proposed from the model-studies, is observed, and related $t_{2g}$- and $e_g$-induced peak positions are moved especially at around the Fermi level (see Supplementary Fig. 5(b,c)). The band widening from the inter-site interaction $V$ increases the overall $d$-$p$ hybridization and changes the bonding- and anti-bonding levels (see the arrows in Fig. 5(b)). At the higher energies beyond 2 eV, an overall pushing-up of the levels can be observed, as in the case of $SrTiO_3$, and at this ranges, $V$ is again acting like an additional $U$. This demonstrates the multifaced role of $V$ in the actual systems.

Comparison with photoemission spectra further signals the apparent role of the long-range interactions. Inclusion of $V$ moves the $e_g$-orbitals away from the Fermi energy, and $e_g$-induced peak (red arrow in Fig. 5(b) and see Supplementary Fig. 5(b)) at around 2.5 eV above $E_F$ is well-separated from the $t_{2g}$-orbitals[37,66,67]. The band widening from the inter-site interaction moves the lower $t_{2g}$-peak (blue arrow in Fig. 5(b)) to -1.5 eV, which is well-matched with the previous spectroscopic studies[68,69] (Supplementary Note 2 and Supplementary Fig. 6). Note that such spectral features can be obtained from GW+DMFT calculations[37], but not from the LDA+DMFT (with local $U$) and GW-only calculations[70,71]. And, here, we access them with a much cheaper computational cost within DFT+$U$+$V$.

For the $d^3$ system, $SrMnO_3$, which has a nominal half-filling in the $t_{2g}$-shell, the main $t_{2g}$ peaks are well-aligned at around the Fermi level for all DFT, DFT+$U$ and DFT+$U$+$V$ calculations (Fig. 5(d)). And this is similar for $d^4$ and $d^5$ cases. While the inclusion of $V$ widens the $d$-bands in the low-energy sector as before, from $d^3$ to $d^5$ systems, the effect is opposite for the higher energy regime and there is small but overall narrowing of the bandwidth (see Fig. 5 and Supplementary Fig. 7). This is because, unlike the cases of $d^1$ and $d^2$, the O-$p$ orbitals are now moved up and located very close to the Fermi level, and the strong $d$-$p$ hybridization deviates the system from the simple $d$-only low-energy models. One can clearly see the peaks that are related to O-$p$ hybridization below 2 eV relative to the Fermi level (see Supplementary Fig. 5). Interestingly, for the later TMOs, $SrMnO_3$, $SrFeO_3$ and $SrCoO_3$, transfer of the weight at the Fermi level peak to higher and lower sector can be seen, and this is reminiscent of the spectral weight transfer from the many-body perspective. Here, this phenomenon arise from the $pd$-hybridization.

For the case of SrCoO$_3$, which has an almost fully occupied $t_{2g}$ shell, the role of $V$ is minimal. The on-site $U$ already redistributes the weights at around the Fermi energy. Further inclusion of $V$ slightly intensifies the spectral weight transfer of $t_{2g}$-peaks to lower band (around -3 eV) without changing the overall electronic structure (see Supplementary Fig. 5(f)). Here we note that while the inclusion of $V$ induces band widening and spectral weight transfers in the Sr$M$O$_3$ series, it does not significantly alter the $d$-electron occupation compared to DFT+$U$ (see Supplementary Fig. 8).

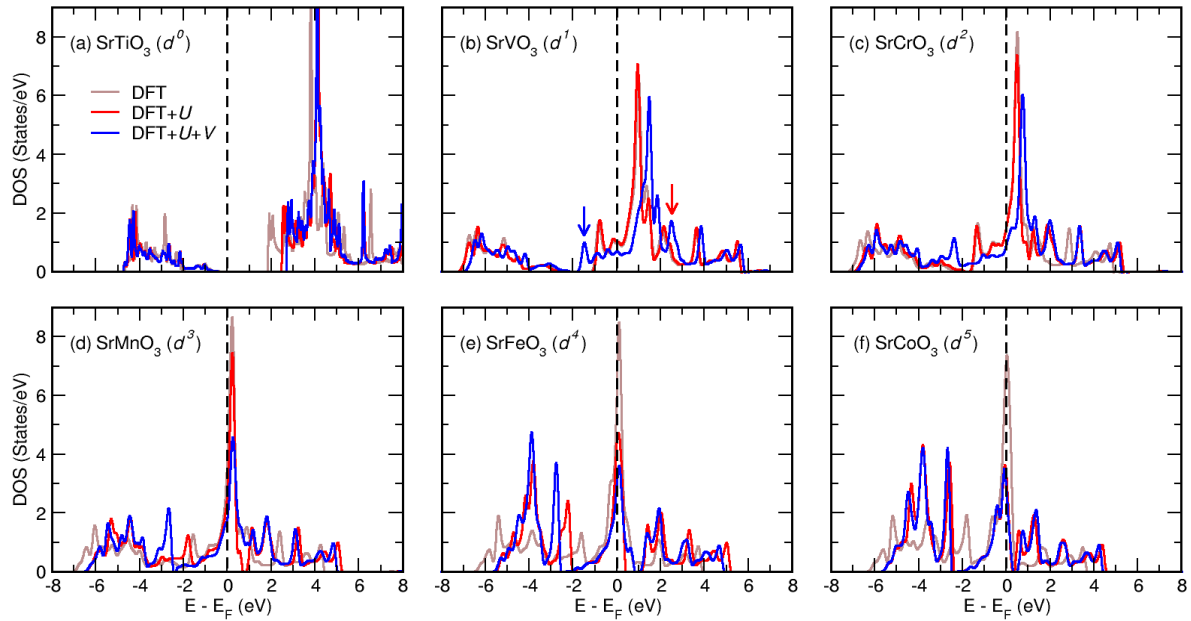

**Fig. 5**. The nonmagnetic $d$-DOS for Sr$M$O$_3$ ($M$ = Ti – Co) are computed at three levels of theory, DFT (solid brown), DFT+$U$ (solid red) and DFT+$U$+$V$ (solid blue). The Fermi level (E$_F$) is set at 0 eV. The red and blue arrows in (b) indicates the bonding- and anti-bonding levels.

### 2-2. Magnetic structures

To demonstrate the role of the long-range Coulomb interaction and its predictability on the materials properties, here, we performed the energetics calculation of the various magnetic configurations for the Sr$M$O$_3$ systems employing DFT, DFT+$U$, and DFT+$U$+$V$. As before, we adopted our cRPA calculated Hubbard $U$ and $V$ parameters from the $d$-$dp$ scheme. The correct estimation of magnetic configuration from DFT-based approach is not a trivial work and small variation of the *onsite* parameter easily switches the stability of different magnetic configurations[72–75]. For our considered Sr$M$O$_3$ series, each has different magnetic ground state, and, here, we considered four different types of collinear configurations: ferromagnetic (FM) and three types of antiferromagnetic (A-AFM, C-AFM, and G-AFM) ones (see Fig. 6(a)).

Figures 6(b)-(f) shows the energetics of different magnetic configurations for all Sr$M$O$_3$ series. Other than the paramagnetism and helical order for SrVO$_3$ and SrFeO$_3$, the experimental magnetic ground states are successfully described from DFT+$U$+$V$ calculations. Especially for SrCrO$_3$ and SrCoO$_3$, where the DFT+$U$ scheme wrongly find the ground phase as A-AFM for both systems, DFT+$U$+$V$ correctly determines the magnetic ground states as C-AFM and FM, respectively. SrCrO$_3$ is a very tricky system to deal with due to various physics involved such as strong correlation, Jahn-Teller distortion, and charge/orbital order[76–79]. While simple DFT correctly describes the correct magnetic order, inclusion of $U$ makes the different magnetic configurations into close energy scales, and A-AFM is selected as ground with meV energy scales. Within the DFT+$U$ scheme, sizable structural distortion, which is larger than the experimental one, is required to obtain the correct C-AFM insulating phase[76,77]. Moreover, the ground state is highly dependent on the local $U$ values[77]. We show that the inclusion of $V$ can successfully stabilizes the C-AFM without resorting to the structural distortion. As the detailed magnetic exchanges, hence the magnetic configurations, depend on the exact electronic structures[80,81], small modification from the long-range interaction is found to be essential here. Similarly, for SrCoO$_3$, where DFT+$U$ fails to predict the correct magnetic ground and, consideration of $V$ accurately finds the experimentally reported FM magnetic ground[72,82,83]. The detailed changes of the magnetic exchanges between TM sites upon the modification of the electronic structures is for the future studies.

For SrVO$_3$, SrMnO$_3$ and SrFeO$_3$, DFT+$U$ and DFT+$U$+$V$ give the same magnetic ground phase. SrVO$_3$ has a paramagnetism, which is not captured from the static calculations. Inclusion of dynamical correlation is needed and, while DFT converges to the nonmagnetic ground, both DFT+$U$ and DFT+$U$+$V$ describe the system as ferromagnetic. SrMnO$_3$ is obtained to be in G-AFM phases, where both DFT+$U$ and DFT+$U$+$V$ present similar energetics for all magnetic configurations. For SrFeO$_3$, our calculations predict a FM ground state, which can be the closest configuration to the experimental magnetic propagation vector with small **q** (0.13, 0.13, 0.13)[84].

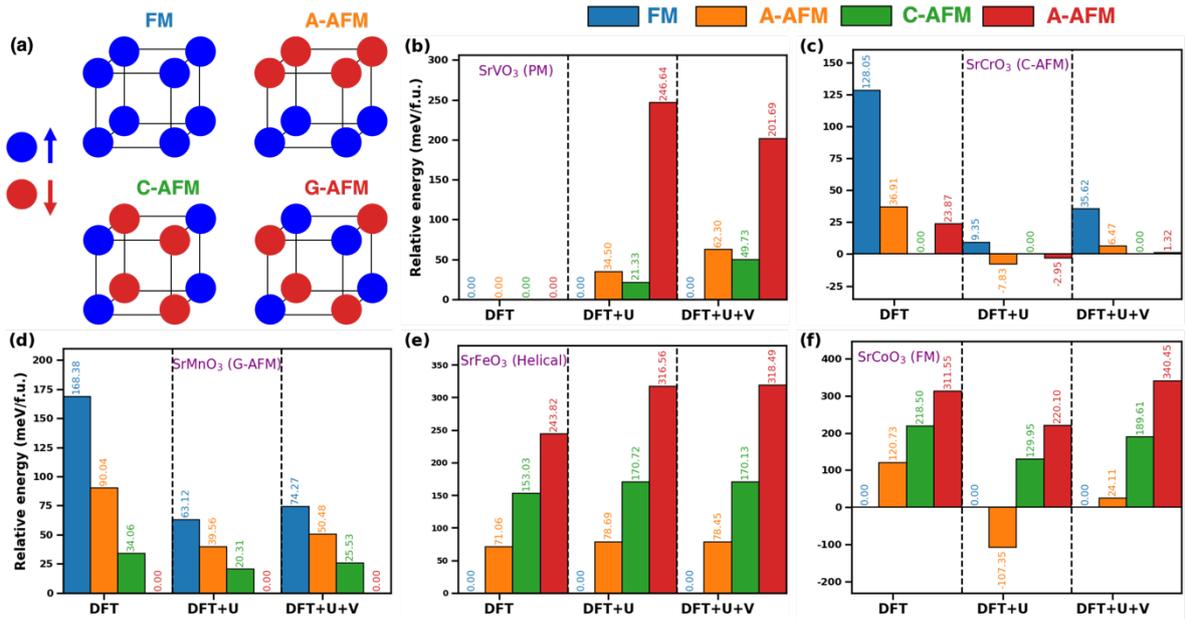

Fig. 6. (a) Schematic representation of four types of collinear magnetic orders: FM, A-AFM, C-AFM, and G-AFM. Transition metal atoms with spin-up (spin-down) alignments are shown as blue (red) spheres. (b-f) Total energy difference per formula unit for different magnetic orders in Sr$M$O$_3$ ($M$ = V – Co) perovskites. The calculations were performed at three different levels of theory: DFT, DFT+$U$, and DFT+$U$+$V$. The total energy difference was calculated with respect to the ground state magnetic configuration energy. The experimental magnetic ordering is given in parentheses.

Before we conclude, we briefly compare our results on the TMO series with previous DFT+$U$+$V$ approaches where the inter-site $V$ is from the TM-$d$ and O-$p$ channels, $V_{pd}$[85–88]. When $V_{pd}$ is included, the local energy levels set by the DFT+$U$ is modified. That is, $V_{pd}$ acts as an additional charge transfer and crystal-field levels, by changing the relative energy levels of O-$p$. For SrTiO$_3$, DFT+$U$+$V_{pd}$ increases the gap over DFT+$U$, but by shifting the *both* O-$p$ and Ti-$d$ levels[85]. This is different from our DFT+$U$+$V$ case where most of the electronic structure changes are from the correlated Ti-$d$ states. Representative features from the long-range correlations such as band renormalization, discussed from the extended Hubbard model[23,24], can only be captured from the DFT+$U$+$V$. As the $d$-occupancy increases, enhanced hybridization between TM-$d$ and O-$p$ states make it difficult to separate out the role of different inter-site interactions. Some features, such as the narrowing of the unoccupied bands[88], and increases band gap with the overall widening at higher energy ranges[87] from the $V_{pd}$ scheme is, in fact, reflection of the $V$ through the hybridized $p$-$d$ channel. We believe further systematic studies are needed to understand the details of the two different approaches.

# Conclusion

In summary, we have evaluated on-site and inter-site Coulomb interactions and have systematically studied the role of the inter-site Coulomb interaction, $V$, on the electronic structures of the representative transition metal oxide perovskites. The screened interaction parameters are calculated from the cRPA calculations. The localization, which increases upon the occupation, competes with the screening effect from various channels, and the resulting interaction parameters show nonmonotonicity upon $d$ orbital occupation. Our results show the nonlocality of the realistic Coulomb interactions and fairly large size of the inter-site interaction, $V$, which can be compared to the local one $U$. The DFT+$U$+$V$ calculations exhibit representative features of the extended Hubbard models such as band renormalization was identified, but as occupation increases enhanced hybridization between TM-$d$ and O-$p$ drives the physics of the system strongly away from the ideal extended Hubbard models. We demonstrate that the inclusion of the inter-site $V$ is essential for the correct reproduction of the experimental magnetic order for Sr$M$O$_3$ systems. We believe that our work can provide guidance for further studies on the real material application of the first-principles theory in the strongly correlated physics community. Especially, 4$d$ and 5$d$ systems with more extended orbitals will be further tackled within the realistic calculation with extended Coulomb interactions[89].

# Methodology

## Calculation of interaction parameters

There are several approaches to obtain the Coulomb interaction parameters from the first-principles electronic structure calculations[90–95]. Here, we adopted the cRPA method, which shares the same spirit with the low-energy model Hamiltonian approaches. In the cRPA scheme, the first step involves constructing the correlated subspace using tools such as maximally localized Wannier functions derived from DFT Kohn-Sham orbitals. The contribution of screening within the correlated target bands, $P^c$, is removed from the total polarizability ($P$). This removal process yields the rest space polarization, $P^r = P - P^c$. Subsequently, the partially screened Coulomb interaction kernel can be obtained by solving the following Bethe-Salpeter type equation:

$$U^{-1} = [U^{bare}]^{-1} - P^r,$$

where $U^{bare}$ and $U$ represent the bare and screened Coulomb interactions, respectively. Then, the frequency ($\omega$)-dependent interaction matrix elements are evaluated among the related Wannier orbitals in the subspace by using the following equation:

$$\boldsymbol{U}_{ijkl}(\omega) = \iint d^3 d^3r' \mathcal{W}_i^*(\boldsymbol{r})\mathcal{W}_k^*(\boldsymbol{r'})U(\boldsymbol{r},\boldsymbol{r'},\omega)\mathcal{W}_j(\boldsymbol{r})\mathcal{W}_l(\boldsymbol{r'})$$

Where $\mathcal{W}(r)$, the target correlated states, are the maximally localized Wannier functions. Here, we employed Wannier functions from the nonmagnetic DFT calculations to exclude additional localization effects from the spin-polarization[96]. For the calculation of the on-site $U$, all target orbitals $\mathcal{W}(r)$ are from the same atom. For the case of $V$, $\mathcal{W}_i(r)$ and $\mathcal{W}_k(r)$ are from one atom, while $\mathcal{W}_j(r)$ and $\mathcal{W}_l(r)$ are from another atom.

Not like the model studies, where the isolated manifolds are considered, in realistic systems, correlated target bands are hybridized with other bands. In our studied systems, Sr$M$O$_3$, the correlated $d$-bands from the transition metals hybridize with the O-$p$ bands, and to obtain the quantitative Coulomb interaction parameters of the $d$-bands, one need to separate out the target bands. Among various schemes, we employed the "projector method" to calculate $P^c$, where the contribution from the target bands is separated out from the Bloch states. The $P^c$ is defined as

$$P^C(\boldsymbol{r},\boldsymbol{r'};\omega) = \sum_{mm'} \psi_m(\boldsymbol{r})\psi_{m'}^*(\boldsymbol{r})\psi_m^*(\boldsymbol{r'})\psi_{m'}(\boldsymbol{r'}) \times \{\frac{1}{\omega - (\epsilon_{m'} - \epsilon_m) + i\eta} - \frac{1}{\omega + (\epsilon_{m'} - \epsilon_m) - i\eta}\}$$

where $\epsilon_m$ and $\eta$ are the Kohn-Sham eigenvalues and a positive infinitesimal, respectively. $|\psi_m\rangle$ is the correlated Bloch function projected onto each Bloch state $|\psi_n\rangle$, with

$$|\psi_m\rangle = \sum_n P_{nm}|\psi_n\rangle$$

where $P_{nm}$ is the projection matrix with the correlated projector defined as

$$P_{nm} = \sum_{i \in C} T_{in}^* T_{im}$$

Here, $i$ indexes the states in the correlated subspace $C$, and $T$ is the unitary transformation matrix between Bloch ($\psi$) and Wannier ($\mathcal{W}$) states, $|\mathcal{W}_i\rangle = \sum_m T_{im}|\psi_m\rangle$. The detailed technicalities of the cRPA based on the projector method can be found in ref. [97]. We have converged our interaction parameters using an 8×8×8 $k$-point sampling, which is in good agreement with the results from the band method[90,98,99]. Eventually, this approach takes account of the hybridization with the *p*-bands, defined as *d-dp* model, such that the Wannier functions are constructed within the whole *d-p* energy windows and screened interactions are calculated in the static limit ($\omega = 0$) by removing the *d*-bands in the polarization calculation. To identify the roles of the *d-p* hybridization, we also considered *d-d* model, where the Wannier functions are constructed only for the *d*-bands and removal of the internal polarization is only within the *d*-bands.

The Wannier functions are obtained using the WANNIER90[100] and the VASP2WANNIER interface[101]. We employed the default parameters for both disentanglement convergence and Wannierization. Specifically, the convergence criterion for disentanglement was set to a relative change of less than $10^{-10} \text{Å}^2$ in the gauge-invariant part of the spread. For Wannierization (localization), the criterion was set to an absolute change of less than $10^{-10} \text{Å}^2$ in the total spread. The employed disentanglement (outer) energy window $\mathbb{W}$ for constructing the Wannier functions is given in Supplementary Table 4. The computational details of the cRPA calculations, as well as the DFT+*U*+*V* for the electronic structure and magnetism calculations are provided in the Supplementary Information (SI), i.e. Supplementary Note 1.

## Acknowledgement

We acknowledge helpful discussions with A. Toschi, G. Sangiovanni, and M. Kim. The authors acknowledge the support from the Advanced Study Group program from PCS-IBS and the hospitality at APCTP, where part of this work was done. B.K. acknowledges support from NRF Grants No. 2021R1C1C1007017, No. 2021R1A4A1031920, and No. 2022M3H4A1A04074153 and the KISTI Supercomputing Center (Project No. KSC-2022-CRE-0465). C.-J.K. was supported by the NRF Grant No. 2022R1C1C1008200 and the KISTI Supercomputing Center (Project No. KSC-2024-CRE-0050). S.K. acknowledges the support of NRF grants (No. 2022R1F1A1063011 and RS-2023-00301914 (G-LAMP Program)).